\documentclass[10pt,twocolumn,showpacs,preprintnumbers,amsmath,amssymb,aps,prl,superscriptaddress,longbibliography]{revtex4-2}
\usepackage{appendix}
\usepackage{bbm}
\usepackage{mathrsfs}
\usepackage{graphicx}
\usepackage{dcolumn}
\usepackage{bm}
\usepackage{braket}
\usepackage{amsmath}
\usepackage{amsfonts}
\usepackage{CJKutf8}
\usepackage{multirow}
\usepackage[dvipsnames]{xcolor}
\usepackage[colorlinks=true,linkcolor=Blue,urlcolor=BlueViolet,citecolor=BlueViolet]{hyperref}
\usepackage{natbib}
\usepackage{floatrow}

\begin{document}

\title{Revisiting quadratic band crossing: from interaction-driven instability to intrinsic topology}
\author{Yadong Jiang}
\affiliation{State Key Laboratory of Surface Physics and Department of Physics, Fudan University, Shanghai 200433, China}
\affiliation{Shanghai Research Center for Quantum Sciences, Shanghai 201315, China}
\author{Linghao Huang}
\affiliation{State Key Laboratory of Surface Physics and Department of Physics, Fudan University, Shanghai 200433, China}
\affiliation{Shanghai Research Center for Quantum Sciences, Shanghai 201315, China}
\author{Zhaochen Liu}
\affiliation{State Key Laboratory of Surface Physics and Department of Physics, Fudan University, Shanghai 200433, China}
\affiliation{Shanghai Research Center for Quantum Sciences, Shanghai 201315, China}
\author{Huan Wang}
\affiliation{State Key Laboratory of Surface Physics and Department of Physics, Fudan University, Shanghai 200433, China}
\affiliation{Shanghai Research Center for Quantum Sciences, Shanghai 201315, China}
\author{Jing Wang}
\thanks{wjingphys@fudan.edu.cn}
\affiliation{State Key Laboratory of Surface Physics and Department of Physics, Fudan University, Shanghai 200433, China}
\affiliation{Shanghai Research Center for Quantum Sciences, Shanghai 201315, China}
\affiliation{Institute for Nanoelectronic Devices and Quantum Computing, Fudan University, Shanghai 200433, China}
\affiliation{Hefei National Laboratory, Hefei 230088, China}

\begin{abstract}
The realization of robust quantum anomalous Hall (QAH) phases at elevated temperatures remains a central challenge in condensed matter physics. While quadratic band crossing points (QBCP) provide a promising route towards QAH states, existing proposals are largely confined to idealized models or hindered by interaction-driven competing orders. Here, we demonstrate that these limitations are not intrinsic to QBCP but arise from their specific implementation. We propose a general mechanism where band inversion between a symmetry-protected orbital doublet (e.g. $d_{xz},d_{yz}$) and an isolated orbital (e.g. $d_{z^2}$)-generically generates a QBCP with opposite curvature. This crossing is directly gapped at the single-particle level by intrinsic atomic spin-orbit coupling, while the underlying band inversion naturally shields the resulting topological gap against other interaction-driven instabilities. We further suggest monolayer compounds $\textit{MNX}_2$ ($\textit{M}$= Ni, Pd, Pt; $\textit{N}$= Nb, Ta; $\textit{X}$= S, Se, Te) as a realistic material class that intrinsically realizes this mechanism. These findings provide a concrete pathway toward robust QAH phases in correlated materials.
\end{abstract}
\maketitle

\emph{Introduction---}The quantum anomalous Hall (QAH) insulator~\cite{chang2023,tokura2019,wang2017c}, characterized by a topologically nontrivial bulk and gapless chiral edge states~\cite{thouless1982,haldane1988,halperin1982}, provides a paradigmatic platform to explore topological quantum matter and holds potential for dissipationless electronic applications~\cite{zhang2012,wang2013a,okazaki2022,patel2024,huang2025}. Despite extensive efforts, experimental realizations of the QAH effect remain restricted to liquid-helium temperatures~\cite{chang2013b,chang2015,mogi2015,bestwick2015,watanabe2019,deng2020,li2024,lian2025,serlin2020,li2021,park2023,xu2023,lu2024,han2024,sha2024,uday2024induced}. This limitation originates from the intrinsic constraints of existing platforms, including magnetic inhomogeneity in magnetic topological insulators~\cite{yu2010,zhang2019,li2019,otrokov2019a,chong2020,garnica2022} and the small energy scales inherent in moir\'e systems~\cite{li2021,park2023,xu2023}. Therefore, achieving robust QAH phases~\cite{you2019,sunj2020,liy2020,Sui2020FeX3,sun2020,li2022,xuan2022,Xu2024PRBai,Jiang2024VWS,yao2024,Jiang2026FeTaX2} at elevated temperatures remains a central challenge in the research of topological materials, motivating the search for intrinsic mechanisms that generate sizable topological gaps without fine tuning.

A quadratic band crossing point (QBCP) in the two-dimensional (2D) Brillouin zone protected by crystalline symmetries has long been proposed as promising starting points for QAH phases~\cite{SunkaiQBT2009,Wu2016QBCPED,Zeng2018QBCPDMRG,liang2017,Rachel2018review,Sur2018,Murray2014RG,Tsai2015Lieb,Liu2025QMC,JiPRBCrCl2,wu2022unprotected}. In principle, its finite density of states enhances interaction effects, allowing spontaneous symmetry breaking and QAH state. However, such QBCP-based routes have remained largely theoretical. A key obstacle is that the topological gap typically arises from interaction-driven symmetry breaking~\cite{SunkaiQBT2009}, which competes with other instabilities such as charge ordering, thereby severely limiting the robustness of the QAH phase in realistic materials. Furthermore, QBCP with the opposite sign of band curvatures—required to stabilize a QAH gap—are exceedingly rare in 2D materials~\cite{wu2022unprotected,JiPRBCrCl2,BGBand2006,BGBand2008,Zhu2016KagomeQBT,Bolens2019Kagome}. 

In this paper, we show that these limitations are not intrinsic to QBCP, but instead arise from how they are typically engineered. We propose a general mechanism in which band inversion in a QBCP system stabilizes the QAH phase against competing instabilities arising from electron interactions. Specifically, we introduce a minimal three-orbital framework consisting of a degenerate orbital doublet and an isolated orbital (e.g. $d_{xz},d_{yz}$ and $d_{z^2}$), whose band inversion generically produces a QBCP with opposite sign of band curvatures. Such a QBCP is directly gapped by intrinsic atomic spin–orbit coupling (SOC), leading to a QAH state, while the underlying band inversion naturally protects the topological gap against other interaction-driven instabilities. Finally, based on density functional theory (DFT) calculations, we suggest a family of compounds \textit{MNX}$_2$ (\textit{M}= Ni, Pd, Pt; \textit{N}= Nb, Ta; \textit{X}= S, Se, Te) as concrete QAH insulator candidates with sizable gaps realized through this mechanism. This establishes a realistic pathway toward robust QAH phases in correlated materials.

\emph{General lattice model---}To capture the essential physics of orbital-driven topological phases, we construct a minimal three-band model on the  tetragonal lattice with $C_{4v}$ symmetry. Each site has three polarized orbitals consisting of a symmetry-protected doublet $|1\rangle,|2\rangle$ with angular momentum ${\ell_z}=\pm 1$ (e.g., $d_{xz},d_{yz}$) and a non-degenerate orbital $|3\rangle$ (e.g., $d_{z^2}$). This configuration is ubiquitous in 2D transition-metal compounds where the effects of the crystal field isolate the $d_{z^2}$ orbital from the $d_{xz},d_{yz}$ manifolds. In momentum space, the tight-binding Hamiltonian is given by:
\begin{eqnarray}\label{model}
&&\mathcal{H}_0(\mathbf{k})
= h_0 + h(\mathbf{k}) 
\nonumber
\\
&&=
\begin{pmatrix}
\mu_1 & i\lambda_a & 0 \\
-i\lambda_a & \mu_1 & 0 \\
0 & 0 & \mu_3
\end{pmatrix}
+
\begin{pmatrix}
h_1(\mathbf{k}) & 0 & h_c(\mathbf{k}) \\
0 & h_1'(\mathbf{k}) & h_c'(\mathbf{k}) \\
h_c^*(\mathbf{k}) & h_c'^*(\mathbf{k}) & h_3(\mathbf{k})
\end{pmatrix},
\end{eqnarray}
where $\mu_{1,3}$ are onsite energies; $\lambda_a$ represents the intrinsic atomic SOC within the doublet, $h_{1,3},h'_{1,3}$ and $h_c,h_c'$ describe intra- and inter-orbital hopping, respectively. Such a description corresponds to a fully spin-polarized ferromagnetic state in realistic materials. The explicit forms of the hopping terms for the tetragonal lattice are summarized in Table~\ref{tab1}.

\begin{table}[b]
\begin{center}
\renewcommand{\arraystretch}{1.5}
\begin{ruledtabular}
\begin{tabular}{lcc}
 & $h^{\rm{Tetr}}(\mathbf{k})$& $h^{\rm{Tri}}(\mathbf{k})$  \\
\hline
$h_{1}$
& $t_{1}\cos k_x+t_{1}'\cos k_y$
& \multirow{2}{*}{$t_{1}\sum\limits_{j=1}^{3}\cos(\mathbf{k}\!\cdot\!\boldsymbol{\delta}_j)$}
\\
$h_{1}'$
& $t_{1}\cos k_y+t_{1}'\cos k_x$
& 
\\
$h_3$
& $t_3 \left(\cos{k_x}+ \cos{k_y}\right)$
& $t_3\sum\limits_{j=1}^{3}\cos(\mathbf{k}\!\cdot\!\boldsymbol{\delta}_j)$
\\
$h_c$
& $i t_c\sin k_x$
&$it_c\big[\sin(\mathbf{k}\!\cdot\!\boldsymbol{\delta}_1)-\frac{1}{2}\sum\limits_{j=2}^{3}\sin(\mathbf{k}\!\cdot\!\boldsymbol{\delta}_j)\big]$
\\
$h_c'$
& $-i t_c\sin k_y$
& 
$-i\frac{\sqrt3}{2} t_c\big[\sin(\mathbf{k}\!\cdot\!\boldsymbol{\delta}_2)-\sin(\mathbf{k}\!\cdot\!\boldsymbol{\delta}_3)\big]$
\\
\end{tabular}
\caption{Explicit forms of $h(\bm{k})$ in the tight-binding model for tetragonal ($h^{\text{Tetr}}(\mathbf{k})$ with $C_{4v}$ symmetry) and trigonal ($h^{\text{Tri}}(\mathbf{k})$ with $C_{6v}$ symmetry) lattices. The orbital basis for both $h(\mathbf{k})$ is $\{d_{xz},d_{yz},d_{z^2}\}$. Here only the nearest neighbor hopping is included. For the trigonal lattice, $\boldsymbol{\delta}_1=(1,0)$, $\boldsymbol{\delta}_2=(-1/2,\sqrt{3}/2)$, and $\boldsymbol{\delta}_3=-(1/2,\sqrt{3}/2)$.}
\label{tab1}
\end{ruledtabular}
\end{center}
\end{table}

\begin{figure}[t]
\begin{center}
\includegraphics[width=3.4in, clip=true]{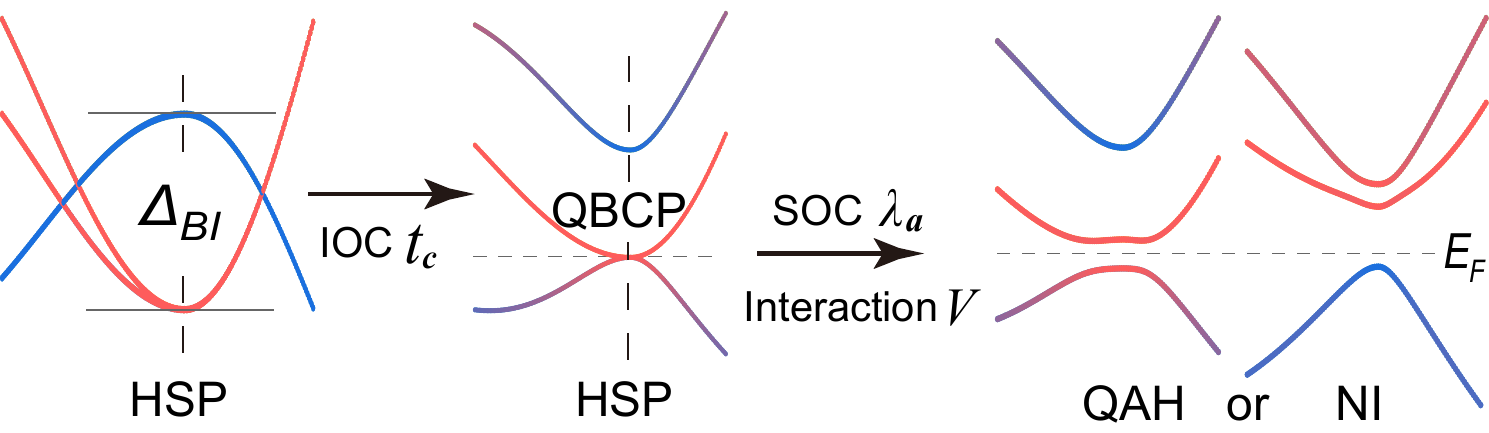}
\end{center}
\caption{Schematic illustration of band inversion generating a QBCP at a high-symmetry point (HSP) in a minimal three-band system. $\Delta_{\rm BI}$ denotes the strength of the band inversion, and $t_c$ the inter-orbital coupling (IOC). The QBCP can be gapped either by atomic spin–orbit coupling $\lambda_a$, yielding a quantum anomalous Hall (QAH) phase, or by electron interaction $V$, leading to either a QAH phase or a normal insulating (NI) phase.}
\label{fig1}
\end{figure} 

The key result of the above model is the emergence of a QBCP with opposite sign of band curvatures. As illustrated in Fig.~\ref{fig1}, the combined effect of band inversion and inter-orbital coupling $t_c$ arising from orbital hybridization leads to a QBCP at high-symmetry points, which is enforced by the $C_{4v
}$ symmetry in the absence of SOC. Specifically, for $h^{\text{Tetr}}(\mathbf{k})$ in Table~\ref{tab1}, the band inversion occurs at $\Gamma=(0,0)$ or $\text{M}=(\pi,\pi)$ when the hopping signs satisfy $\mathrm{sgn}(t_1)=\mathrm{sgn}(t_1^\prime)\neq \mathrm{sgn}(t_3)$. The strength of this inversion is quantified by:
\begin{equation}
\Delta_{\text{BI}}=\mu_3-\mu_1+ t_1+t_1^\prime-2t_3,
\end{equation}
where a positive $\Delta_{\text{BI}}$ represents $d_{z^2}$ located above the $d_{xz},d_{yz}$ orbitals at the band inversion point. The atomic SOC $\lambda_a$ gaps the QBCP, and the resulting lower band acquires a quantized Berry phase of $\text{sgn}(\lambda_a)2\pi$. When the chemical potential is in the gap (i.e., the lower band is fully occupied), this directly yields a QAH state with Hall conductance $\sigma_{xy}=\mathcal{C}e^2/h$ with Chern number $\mathcal{C}=\text{sgn}(\lambda_a)$.

\emph{Electron interactions and phase stability---}While previous QBCP-based proposals rely on electron interactions to spontaneously break symmetry and generate a topological gap, our model naturally realizes a QAH phase via intrinsic atomic SOC. However, since electron-electron repulsions can induce competing symmetry-breaking orders, it is essential to evaluate the stability of the QAH phase. We incorporate the onsite density-density interactions via the minimal Hamiltonian:
\begin{equation}
    \mathcal{H}_{\rm{int}}=\sum_{i}V_1 n_{i,1}n_{i,2}+V_2(n_{i,1}+n_{i,2})n_{i,3},
    \label{eq:H_int}
\end{equation}
where $n_{i,\alpha}\equiv c^\dagger_{i,\alpha} c_{i,\alpha}$ is the density operator for the $\alpha$ orbital at the $i$-th lattice site, $\alpha=1,2,3$ corresponds to $d_{xz},d_{yz},d_{z^2}$ orbitals, respectively. $V_1$ represents the intra-doublet repulsion (within $d_{xz},d_{yz}$ orbitals) and $V_2$ denotes the interaction between $d_{xz},d_{yz}$ and $d_{z^2}$. For simplicity, we set $V_1=V_2=V$, as this does not qualitatively alter the phase diagram. We employ the self-consistent Hartree-Fock approximation to capture the interplay between the interaction strength $V$ and the band-inversion parameter $\Delta_{\mathrm{BI}}$.

Before presenting the numerical phase diagram, we identify potential symmetry-breaking orders. Beyond the NI and QAH states, the system may host three distinct nematic phases that break $C_{4v}$ symmetry. These phases are characterized by the density matrix $\rho_{\alpha\alpha'}=\langle c^\dagger_{i,\alpha} c_{i,\alpha'}\rangle$, where $\langle\cdots\rangle$ denotes the statistical average:
\begin{itemize}
    \item $\rho_{11} \neq \rho_{22}$, lifting the $d_{xz},d_{yz}$ degeneracy.
    \item $\text{Re}\rho_{12}\neq 0$, mixing the $d_{xz},d_{yz}$ components.
    \item $\rho_{23},\rho_{13}\neq 0$, inducing $\{d_{xz},d_{yz}\}$–$d_{z^2}$ hybridization.
\end{itemize}

\begin{figure}[t]
\begin{center}
\includegraphics[width=3.4in, clip=true]{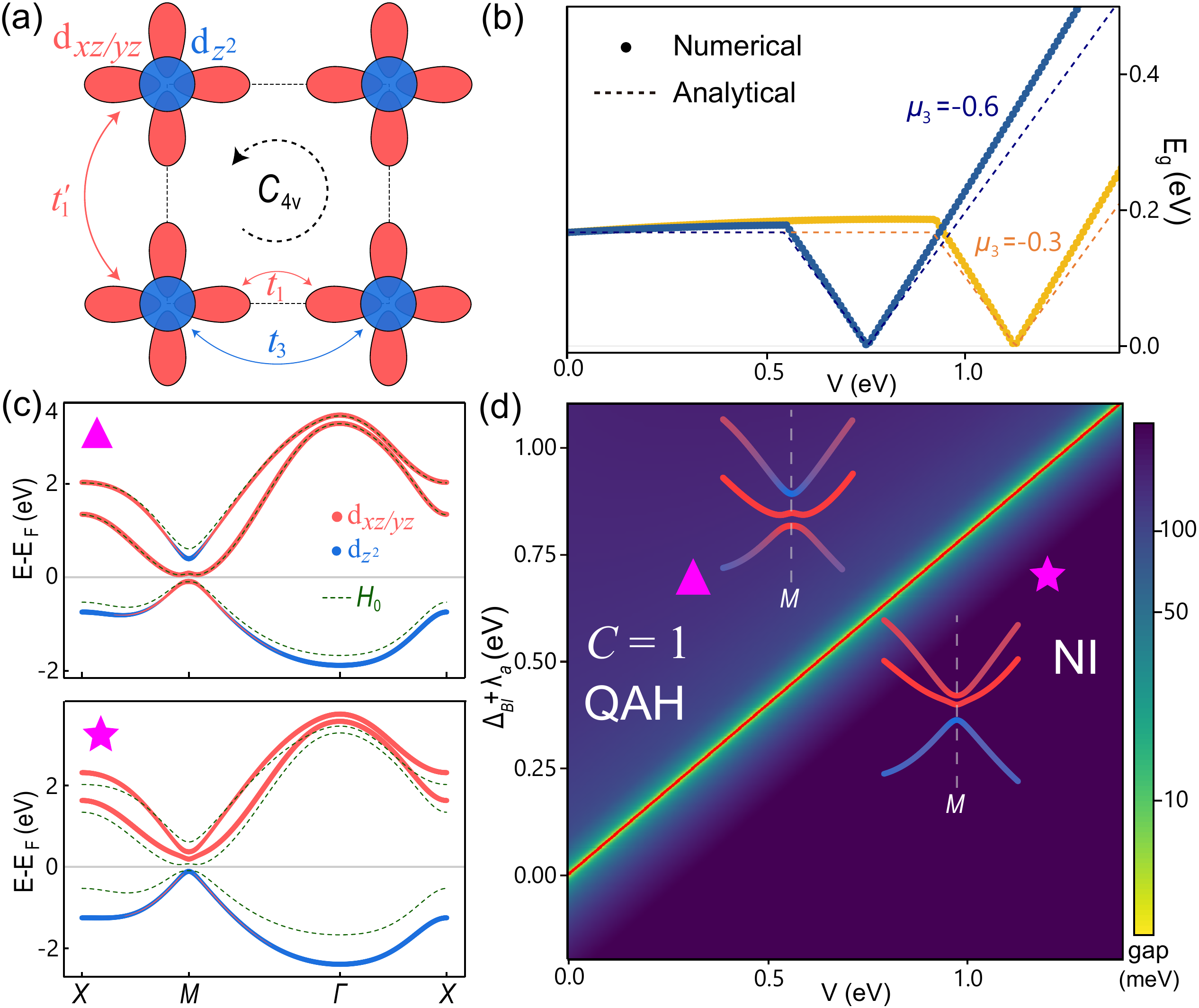}
\end{center}
\caption{Interaction effects and phase stability in the tetragonal lattice. (a) Schematic of the minimal three-band model. The quadratic band crossing is protected by the $C_{4v}$ symmetry without SOC. (b) Topological gap $E_g$ at the M point as a function of the interaction strength $V$. The circles and dashed line represent the numerical and analytical results {\color{blue}\cite{supple}}, respectively. (c) Hartree-Fock band structures (solid lines) compared with the non-interacting model $\mathcal{H}_0$ (dashed lines). Top and bottom panels illustrate the QAH and NI states with the corresponding points in the phase diagram (d). (d) Hartree-Fock phase diagram in the $V$ vs. $(\Delta_{\text{BI}}+\lambda_a)$ plane. The logarithmic color scale indicates the magnitude of the band gap, highlighting the linear phase boundary (red line) between the QAH and NI phases. Parameters (in eV): $\mu_1=1.71$, $t_1=1.01$, $t'_1=0.68$, $t_3=-0.57$, $t_c=1.04$, and $\lambda_a=0.0837$, the band-inversion strength $\Delta_{\text{BI}}$ is tuned via $\mu_3$.}
\label{fig2}
\end{figure}

The Hartree-Fock calculations, summarized in Fig.~\ref{fig2}, demonstrate that the QAH phase is remarkably robust. Crucially, we find no evidence of nematic instabilities within the physically relevant parameter regime. Instead, the system undergoes a direct QAH-to-NI transition. As shown in the orbital-projected band structures [Fig.~\ref{fig2}(c)], increasing the interaction $V$ acts to suppress the band inversion. At a critical interaction strength, the band inversion is completely eliminated, driving the system into the NI phase.

The phase boundary in the $V$-$(\Delta_{\text{BI}}+\lambda_a)$ plane exhibits a striking linear scaling [Fig.~\ref{fig2}(d)], which can be understood through a perturbative analysis. In the limit of small SOC, the hybridization between $d_{z^2}$ and $d_{xz},d_{yz}$ is negligible, except near the band inversion point. Consequently, the density matrix remains nearly diagonal, with the $d_{z^2}$ orbital nearly fully occupied ($\rho_{33}\approx1-2n$) and the $d_{xz},d_{yz}$ orbitals sparsely populated ($\rho_{11}=\rho_{22}\equiv n$). In this regime, the primary effect of the interaction is captured by the Hartree term, $\mathcal{H}_{\rm{int}}^{\rm{HF}}\sim V\text{diag}\{\rho_{22}+\rho_{33},\rho_{11}+\rho_{33},\rho_{11}+\rho_{22}\}$, which acts as an orbital-dependent potential shift. The resulting band energies at the inversion point are: $E_3 = \mu_3-2t_3+2nV,~E_{2,1}=\mu_1-t_1-t'_1\pm\lambda+(1-n)V$. 
The fundamental gap is then defined as  $E_g=\text{min}\{E_2-E_1,E_3-E_1\}$. The gap-closing condition, $E_3=E_1$, yields a linear relationship for the critical interaction strength:
\begin{equation}\label{phase_boundary}
    (1-3n)V=\Delta_{\mathrm{BI}}+\lambda_a.
\end{equation}
This confirms that the critical interaction scales directly with the magnitude of band inversion $\Delta_{\text{BI}}$. Given that the occupation number $n$ is small and only weakly perturbed by $\Delta_{\text{BI}}$ and $V$, this linear boundary is remarkably stable. As shown in Figs.~\ref{fig2}(b) and~\ref{fig2}(d), our analytical fits for the topological gap and the phase boundary (using $n=1/15$) show excellent agreement with the full numerical Hartree-Fock results. Notably, Fig.~\ref{fig2}(b) illustrates that the band inversion between the extremal bands effectively shields the QAH phase from many-body repulsions. In the weak-interaction regime, this protection manifests as a nearly constant topological gap ($\sim2\lambda_a$), underscoring the intrinsic stability of the mechanism. Furthermore, the negligible influence of the exchange (Fock) term indicates that the previously discussed nematic instabilities are energetically suppressed, confirming the robustness of the orbital-driven QAH state.

\begin{figure}[b]
\begin{center}
\includegraphics[width=3.4in, clip=true]{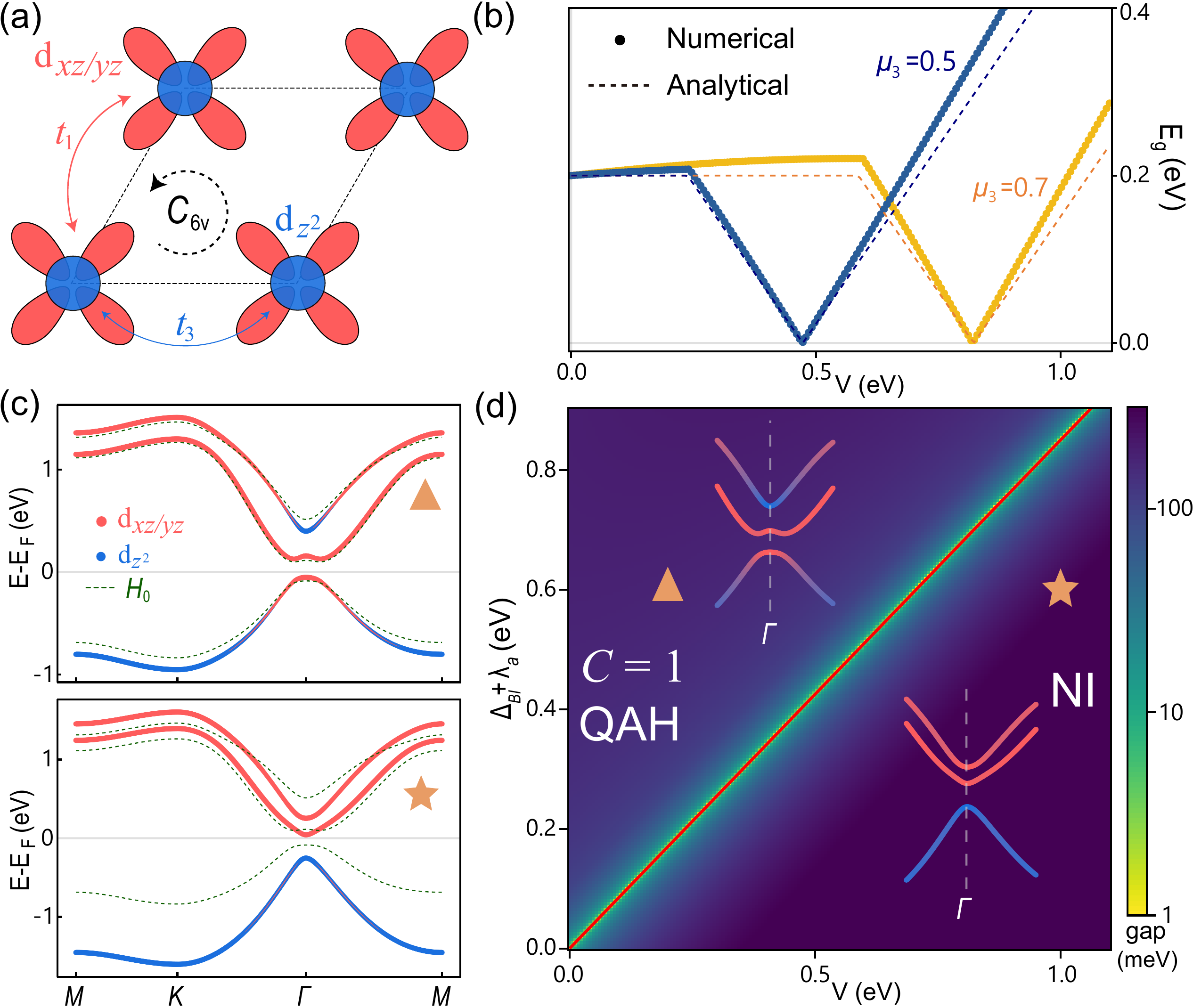}
\end{center}
\caption{Interaction effects and phase stability in the trigonal lattice. (a) Schematic of the three-band model. The quadratic band crossing is protected by the $C_{6v}$ symmetry without SOC. (b) Topological gap $E_g$ at the $\Gamma$ point as a function of the interaction strength $V$. (c) Hartree-Fock band structures (solid lines) compared with the non-interacting model $\mathcal{H}_0$ (dashed lines). Top and bottom panels illustrate the QAH and NI states, respectively, which corresponds to the points in the phase diagram (d). (d) Hartree-Fock phase diagram in the $V$ vs. $(\Delta_{\text{BI}}+\lambda_a)$ plane.  Parameters (in eV): $\mu_1=2.0$, $t_1=-0.3$, $t_3=0.3$, $t_c=0.4$, and $\lambda_a=0.1$, the band-inversion strength $\Delta_{\text{BI}}$ is tuned via $\mu_3$.}
\label{fig3}
\end{figure}

\emph{Extension to $C_{6v}$ system---}To demonstrate the universality of the above mechanism, we extend our analysis to the trigonal lattice with $C_{6v}$ symmetry. The corresponding Hamiltonian $h^{\text{Tri}}(\mathbf{k})$ is constructed using the same minimal three-band framework (see Table~\ref{tab1}). In this high-symmetry setting, a QBCP emerges at the $\Gamma$ point, where it is protected by the $C_{6v}$ symmetry in the absence of SOC. The strength of the band inversion at $\Gamma$ is determined by $\Delta_{\mathrm{BI}}=\mu_3-\mu_1-3 t_1+3 t_3$. As with the tetragonal case, the introduction of atomic SOC gaps the QBCP at the single-particle level, resulting in a QAH conductance $\sigma_{xy}=\mathcal{C}e^2/h$ with a Chern number $\mathcal{C}=\mathrm{sgn}(\lambda_a)$.

We further evaluate the stability of the QAH phase under electron interactions. Self-consistent Hartree-Fock calculations are summarized in Fig.~\ref{fig3}. The phase diagram mirrors the behavior observed in the tetragonal system, with a linear phase boundary governed by the same analytic relation [Eq.~(\ref{phase_boundary})]. For the trigonal lattice, the boundary is accurately described using an effective occupation parameter $n=0.05$. This consistency across distinct symmetries underscores the intrinsic stability of the orbital-driven QAH effect.

\begin{figure}[t]
\begin{center}
\includegraphics[width=3.4in, clip=true]{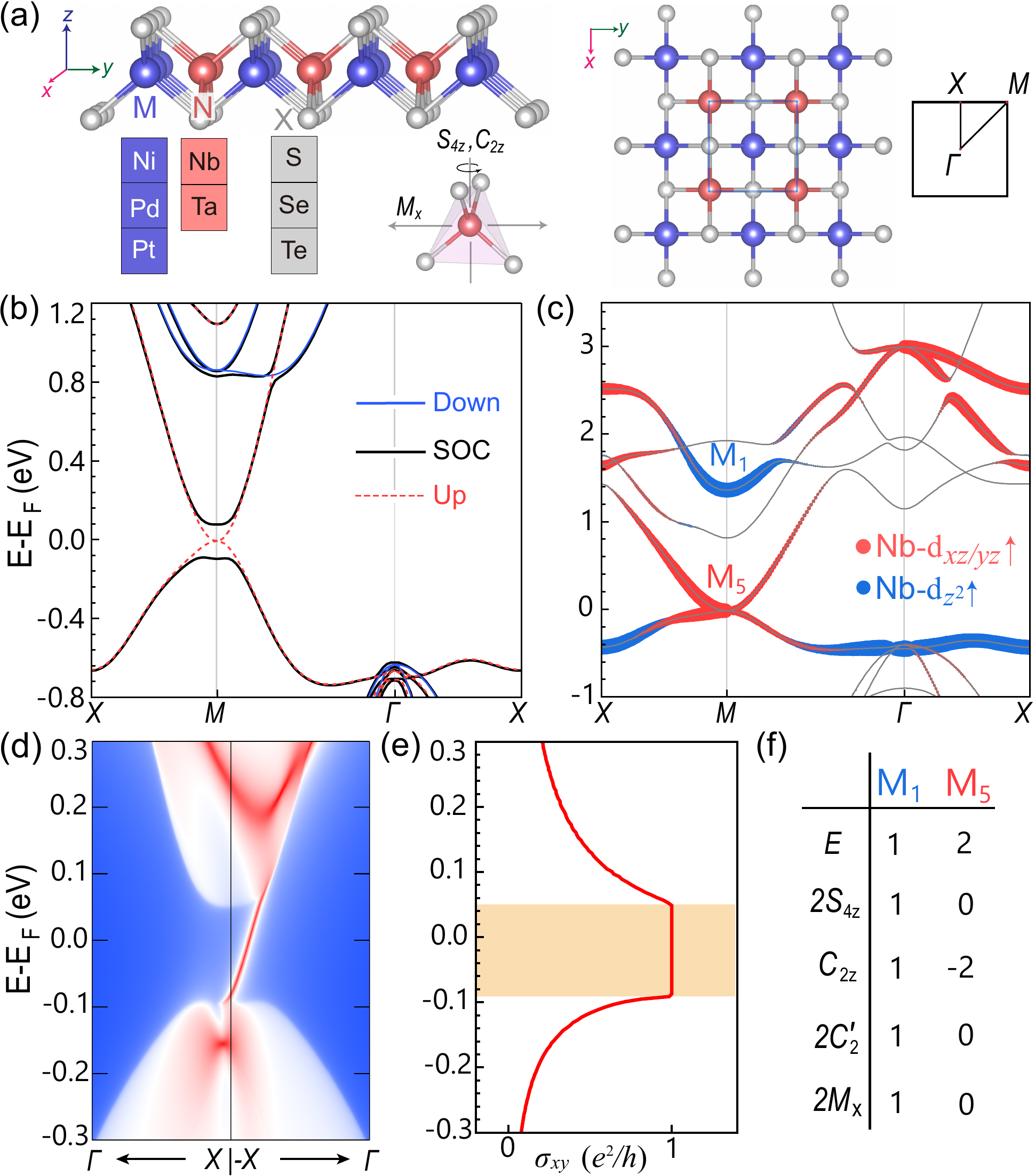}
\end{center}
\caption{(a) Atomic structure of monolayer \textit{MNX}$_2$ family from the side and top views. The key symmetry operations of $P$-$4m2$ include $S_{4z}$, $C_{2z}$ rotations and mirror symmetry $M_x$ ($M_y$), where $S_{4z}\equiv\mathcal{I}C_{4z}$ and $\mathcal{I}$ is inversion symmetry. (b) The band structure of monolayer PdNbSe$_2$ without (blue and red lines) and with (black line) SOC. (c) The $d$-orbital projection band structures without SOC of monolayer PdNbSe$_2$. (d,e) Topological edge states calculated along the $x$ axis; and anomalous Hall conductance $\sigma_{xy}$ as a function of Fermi energy, respectively. The shaded regions in (e) denote the topological gap. (f) Partial character table of little group $D_{2d}$ at M point. The rotation axis of $C_2^\prime$ symmetry is along $(\hat{x}-\hat{y})$ direction.}
\label{fig4}
\end{figure}

\emph{Candidate materials---}We suggest a class of $\textit{MNX}_2$ materials that crystallize in a tetragonal lattice as a concrete realization of the robust QAH mechanism. The monolayer $\textit{MNX}_2$ belongs to the space group $P$-$4m2$ (No.~$115$). As shown in Fig.~\ref{fig4}(a), each primitive cell consists of a trilayer structure where an \textit{MN} atomic layer is sandwiched between two \textit{X} layers. \textit{M} and \textit{N} are transition metals, coordinated by four chalcogens \textit{X},  forming edge-sharing tetrahedra. DFT calculations suggest that the materials listed in Table~\ref{tab2} are candidates for QAH insulators. Structural relaxations and magnetic properties are obtained within DFT+$U$~\cite{kresse1996,perdew1996,dudarev1998}, while electronic structures are confirmed by using the Heyd-Scuseria-Ernzerhof hybrid functional~\cite{hse06}. The fully optimized lattice constants are listed in Table~\ref{tab2}, and the dynamical stability is confirmed by first-principles phonon calculations~\cite{supple}. 

\begin{table}[b]
\caption{Lattice constant $a$ (\AA); Curie temperature $T_c$ (K) from Monte Carlo simulations; band gap $E_g$ (meV).}
\begin{center}\label{tab2}
\renewcommand{\arraystretch}{1.5}
\begin{tabular*}{3.4in}
{@{\extracolsep{\fill}}lccclccc}
\hline
\hline
Materials &$a$ & $T_c$& $E_g$& Materials &$a$& $T_c$& $E_g$ \\
\hline
NiNbS$_2$  & 3.87 & 229 & 163 & PdNbS$_2$  & 4.04 & 256 & 36  \\
NiNbSe$_2$ & 3.91 & 236 & 259 & PdNbSe$_2$ & 4.09 & 300 & 167 \\ 
PtNbS$_2$  & 4.04 & 284 & 323 & PdNbTe$_2$ & 4.19 & 311 & 268 \\
PtNbSe$_2$ & 4.07 & 298 & 338 & NiTaS$_2$  & 3.78 & 66  & 150\\
PtNbTe$_2$ & 4.15 & 277 & 303 & NiTaTe$_2$ & 3.87 & 38  & 100 \\
PtTaS$_2$  & 3.99 & 263 & 197 & PtTaSe$_2$ & 4.00 & 160 & 263 \\
\hline
\hline
\end{tabular*}
\end{center}
\end{table}

First-principles calculations reveal that all compounds listed in Table~\ref{tab2} exhibit ferromagnetic ground state with an out-of-plane easy axis. All candidates share similar characteristics of topological bands, we focus on PdNbSe$_2$ as a representative example. Fig.~\ref{fig4}(b) displays the band structure with and without SOC. $d$-orbital projection band structure in Fig.~\ref{fig4}(c) shows that a spin-polarized QBCP emerges at M point, originating from the band inversion between degenerate Nb-($d_{xz}^\uparrow, d_{yz}^\uparrow$) orbitals (2D irreducible representation M$_5$) and Nb-$d_{z^2}^\uparrow$ orbital (1D irreducible representation M$_1$). The inclusion of SOC opens a non-trivial gap at QBCP, and the calculated anomalous Hall conductance $\sigma_{xy}$ reaches a quantized value of $e^2/h$ within the bulk gap [Fig.~\ref{fig4}(e)]. This confirms a Chern number $\mathcal{C}=1$, consistent with the single chiral edge state observed in the edge local density of states calculations [Fig.~\ref{fig4}(d)]. By mapping these DFT results to our minimal tight-biding model, we find that $h^{\rm{Tetr}}(\mathbf{k})$ faithfully captures the intrinsic topology of the \textit{MNX}$_2$ family. Interestingly, we find that the band topology is remarkably robust within DFT+$U$ calculations under various $U$ values~\cite{supple}, which is consistent with the above general mechanism.

\emph{Discussions---}We have established a general three-band framework where a QBCP emerges as a natural consequence of band inversion between symmetry-protected degenerate orbitals and an auxiliary singlet band. This mechanism is rooted in a ubiquitous orbital configuration common to multi-orbital systems, making the resulting QAH topology a robust outcome of intrinsic atomic SOC. Furthermore, this mechanism can be generalized to high Chern number states. As detailed in the Supplemental Material, we extend our minimal three-band framework to a trigonal lattice using a basis of $\{d_{x^2-y^2},d_{xy},d_{z^2}\}$ orbitals, which realizes a robust $\mathcal{C}=2$ QAH state~\cite{supple}.

A key implication of our framework is its fundamentally distinct stability criterion compared to interaction-driven QBCP proposals. In earlier checkerboard or honeycomb models~\cite{SunkaiQBT2009,Wu2016QBCPED,Zeng2018QBCPDMRG}, the topological gap depends on spontaneous symmetry breaking, which must compete with various charge order instabilities. In our framework, the QAH phase is ``pre-formed'' by atomic SOC. Consequently, electron-electron interactions primarily serve to renormalize the band dispersion and reduce the band inversion parameter $\Delta_{\text{BI}}$. Our Hartree-Fock analysis confirms this resilience, revealing a simplified phase structure consisting only of the QAH and NI phases, without intervening symmetry-broken nematic or ordered states. This highlights that the robustness of QAH phase originates from band inversion itself, rather than from a delicate balance of competing interaction-driven orders.

This robustness offers a clear advantage over other atomic-SOC-induced QAH mechanisms, such as those in monolayer transition metal trihalides (\textit{MX}$_3$, where \textit{M} is a transition metal and \textit{X} is a halide). In those systems, the non-trivial gaps are often fragile, easily collapsed by strong electron-electron interactions that drive the system toward topologically trivial states~\cite{Mellaerts2021Model,Li2019FeX3,Sui2020FeX3,Xu2024PRBai,xu2024}. In our QBCP-based route, the gap is shielded by the band inversion $\Delta_{\mathrm{BI}}$, remaining stable even when the characteristic interaction scale exceeds the SOC strength, a condition often encountered in realistic $d$-orbital systems. 

\begin{acknowledgments}
\emph{Acknowledgments---}We thank Jingjing Gao for helpful discussions. This work is supported by the National Key Research Program of China under Grant No.~2025YFA1411400; Natural Science Foundation of China under Grant No.~12350404; Quantum Science and Technology-National Science and Technology Major Project through Grant No.~2021ZD0302600; and the Science and Technology Commission of Shanghai Municipality under Grants No.~23JC1400600, No.~24LZ1400100 and No.~2019SHZDZX01, and it is sponsored by the ``Shuguang Program'' supported by the Shanghai Education Development Foundation and Shanghai Municipal Education Commission. Y.J. is supported by the China Postdoctoral Science Foundation under Grants No.~GZC20240302 and No.~2024M760488. 

Y.J. and L.H. contributed equally to this work.
\end{acknowledgments}

\end{document}